\documentclass[12pt]{article}

\begin{document}
\begin{center}
{\bf ``Square Root" of the Proca Equation: Spin-3/2 Field Equation}\\
\vspace{5mm}
 S.I. Kruglov \footnote{On leave from National Scientific Centre of Particle and High
 Energy Physics, M. Bogdanovich St. 153, Minsk 220040, Belarus.}\\
\vspace{5mm}
\textit{University of Toronto at Scarborough,\\ Physical and Environmental Sciences Department, \\
1265 Military Trail, Toronto, Ontario, Canada M1C 1A4}
\end{center}

\begin{abstract}
New equations describing particles with spin 3/2 are derived. The
non-local equation with the unique mass can be considered as
``square root" of the Proca equation in the same sense as the
Dirac equation is related to the Klein-Gordon-Fock equation. The
local equation describes spin 3/2 particles with three mass
states. The equations considered involve fields with spin-3/2 and
spin-1/2, i.e. multi-spin 1/2, 3/2. The projection operators
extracting states with definite energy, spin, and spin projections
are obtained. All independent solutions of the local equation are
expressed through projection matrices. The first order
relativistic wave equation in the 20-dimensional matrix form, the
relativistically invariant bilinear form and the corresponding
Lagrangian are given. Two parameters characterizing non-minimal
electromagnetic interactions of fermions are introduced, and the
quantum-mechanical Hamiltonian is found. It is proved that there
is only causal propagation of waves in the approach considered.

\end{abstract}

\section{Introduction}

Until now, we know different field equations for higher spin
particles introduced in \cite{Dirac}, \cite{Fierz}, \cite{Wigner},
\cite{Bargman}, \cite{Bhabha}, \cite{Hurley}. It should be
mentioned that the higher spin (HS) gauge theories are intensively
investigated now \cite{Vasiliev}. In such an approach the higher
derivatives appear in field interactions, so that the theory is
nonlocal. The HS gauge theories include linearized electrodynamics
and gravity. The superstring field theory can be connected to some
version of the HS gauge theory. Anti-de Sitter (AdS) space is a
natural background for interacting HS gauge fields. We also
mention the Maldacena conjecture \cite{Maldacena} that is the
correspondence between field (string) theories on
$(d+1)$-dimensional AdS space and (super) conformal field theories
(CFT) on the boundary ($d$-dimensional) of this space. For
example, super $SU(N)$ Yang-Mills theory ($N=4$) may be formulated
as $IIB$ supergravity compactified on $AdS_5\times S^5$.

The Rarita-Schwinger field of spin-3/2 \cite{Rarita} appears
naturally in the framework of AdS/CFT correspondence as a solution
on the boundary \cite{Volovich}. The supersymmetry breaking in
supergravity results in appearing massive spin-3/2
fields-gravitinos. So, gravitino plays very important role in
supersymmetrical theory of gravity.

It is well known, however, that there are physical inconsistences
for describing massive particles with higher spins $s$, $s\geq 1$
in external electromagnetic fields. So, free higher spin particles
can be described without difficulties, but if one introduces
interactions of particles with electromagnetic fields, some
problems arise \cite{Velo}. For example, in the framework of the
Rarita - Schwinger approach, there is superluminal speed of
particles interacting (minimally) with electromagnetic fields.
This indicates on noncausal propagation of interacting particles
\cite{Zwanziger}. This difficulty, however, can be eliminated by
considering $N=2$ supergravity \cite{Deser}.

For phenomenological applications of hadron physics, one needs a
consistent model with an electromagnetic background. Massive
spin-3/2 electrodynamics was studied in \cite{Deser1}.

The purpose of this paper is to take the ``square root" of the
Proca equation \cite{Proca} in analogous procedure performed by
Dirac \cite{Dirac2} on the Klein-Gordon-Fock equation, and to
obtain the spin-3/2 field equation. We concentrate here on
studding solutions (pure spin states) of free equations and
discussing particle electromagnetic interactions. Possibly,
presented here new equations for spin-3/2 particles can be
considered as effective equations for describing hadrons.

The paper is organized as follows. In Sec. 2 we derive the
``square root" of the Proca equation, which is the spin-3/2
non-local field equation describing particles with the unique
mass. Then, the local spin-3/2 field equation is obtained. The
projection operators extracting solutions of the field equation
with positive and negative energy are constructed in Sec. 3. It
was proved that there are three different mass states of fields.
We consider projection operators in Sec. 4, which allow us to
separate states with spins $1/2$ and $3/2$ as well as spin
projections. It is shown that the field equation considered
describes fields with spin $1/2$ and spin $3/2$, i.e. multi-spin
$1/2$, $3/2$. In Sec. 5 we represent the second order equation in
the 20-dimensional matrix form of the first order relativistic
wave equation. The relativistically invariant bilinear form and
the corresponding Lagrangian are given there. Sec. 6 is devoted to
introducing two parameters characterizing non-minimal
electromagnetic interactions of fermions. We also find the
quantum-mechanical Hamiltonian. Sec. 7 is devoted to discussion.

The system of units $\hbar =c=1$ is used. Four-vectors are defined
as $v^2_\mu=v^2_m +v^2_4 =\textbf{v}^2-v^2_0$, $\textbf{v}^2\equiv
v^2_m=v_1^2 +v_2^2+v_3^2$, $v_4=iv_0$.

\section{Spin-3/2 Field Equation}

We are looking for an equation which when squared will produce the
Proca equation. The Proca equation \cite{Proca} for a free
particle possessing the mass $m$ is given by
\begin{equation}
\partial_\nu \varphi_{\mu\nu}(x) +m^2\varphi_\mu (x) =0 ,  \label{1}
\end{equation}
where $\partial_\nu =\partial/\partial x_\nu =(\partial/\partial
x_m,\partial/\partial (it))$, and antisymmetric tensor
$\varphi_{\mu\nu}(x)$ can be expressed through the four-vector
field $\varphi_\mu (x)$ as follows
\begin{equation}
 \varphi_{\mu\nu} (x)=\partial_\mu\varphi_\nu (x)- \partial_\nu\varphi_\mu
 (x) .  \label{2}
\end{equation}
Replacing the tensor $\varphi_{\mu\nu}(x)$ from Eq. (2) into Eq.
(1), one transfers the system of first order equations (1), (2)
into the second order equation
\begin{equation}
\left[\left(\partial^2_\alpha -m^2\right)\delta_{\mu\nu}
-\partial_\mu\partial_\nu \right]\varphi_\nu (x)=0 . \label{3}
\end{equation}
We can represent Eq. (3) in the matrix form
\begin{equation}
N\varphi (x)=0 , ~~ N =\left(\partial^2_\alpha
-m^2\right)I_4-(\partial \cdot\partial) ,\label{4}
\end{equation}
where the $I_4$ is the unit $4\times 4$ matrix and the $(\partial
\cdot\partial)$ is the matrix-dyad with the matrix elements
$(\partial \cdot\partial)_{\mu\nu}=\partial_\mu\partial_\nu$, so
that the $N$ is the matrix-differential operator.

Let us consider the operator equation
\begin{equation}
\left[\gamma_\mu\partial_\mu+ a\frac{(\partial \cdot
\partial)}{\partial^2_\alpha}+m \right]\left[\gamma_\mu\partial_\mu- a\frac{(\partial \cdot
\partial)}{\partial^2_\alpha}-m  \right]=N
, \label{5}
\end{equation}
where $a$, in general, is an operator, and the Dirac matrices
$\gamma_\mu $ obey the commutation relations (we use notations as
in \cite{Ahieser})
\begin{equation}
\gamma_\mu \gamma_\nu +\gamma_\nu \gamma_\mu =2\delta_{\mu\nu} .
\label{6}
\end{equation}
We notice that Croneker symbols are used in Eqs. (3), (6) instead
of Minkowski tensors because of our Euclidian metrics (fourth
components of all vectors are imaginary). It is easy to check that
Eq. (5) is valid for two values of the $a$:
\begin{equation}
a=-m \pm \sqrt{m^2+\partial_\alpha^2} . \label{7}
\end{equation}
Thus, the operators in Eqs. (5), (7) are nonlocal and act in the
momentum space as
\begin{equation}
1/\partial_\alpha^2 \rightarrow - 1/p^2
,~~~~\sqrt{m^2+\partial_\alpha^2}\rightarrow\sqrt{m^2-p^2}
,\label{8}
\end{equation}
where the four-momentum squared being $p^2=\textbf{p}^2-p_0^2$.
From operator equation (5), we come to nonlocal equations for the
spinor-vector $\psi (x)=\{\psi_\mu (x)\}$:
\begin{equation}
\left[\gamma_\mu\partial_\mu+ a\frac{(\partial \cdot
\partial)}{\partial^2_\alpha}+m \right]\psi (x)=0 ,
 \label{9}
\end{equation}
or
\begin{equation}
\left[\gamma_\mu\partial_\mu- a\frac{(\partial \cdot
\partial)}{\partial^2_\alpha}-m  \right]\psi (x)=0 .
\label{10}
\end{equation}
As in the similar procedure performed by Dirac on the
Klein-Gordon-Fock equation, we have two equations with the
opposite signs of the mass.

For massless case, one has to put $m=0$, $a=\sqrt{\partial_\mu^2}$
in Eqs. (9), (10). So, from Eq. (9), we find that massless
particles are described by a nonlocal equation
\[
\left[\gamma_\mu\partial_\mu+ \frac{(\partial \cdot
\partial)}{\sqrt{\partial_\mu^2}}\right]\psi (x)=0 .
\]

 In further, the attention will be paid on Eq.
(9). The 4-vector-spinor $\psi_\mu (x)$ without supplementary
conditions possesses the $16$ components and describes a field
with spin-$3/2$ ($8$ components, four spin projections $\pm 1/2,
\pm 3/2$, and two values, positive and negative, of energy) and
two copies of spin-$1/2$ fields (four components for each
spin-$1/2$ field). Thus, the 4-vector-spinor $\psi_\mu (x)$, in
general, describes fields with multi-spin $1/2, 3/2$. In
\cite{Kruglov86}, \cite{Kruglov2001}, we considered the
non-Abelian gauge theory of particles with multi-spin $1/2$, $3/2$
described by the 4-vector-spinor $\psi_\mu (x)$ which obeys the
Dirac equation. The equations (9), (10) are different from that by
the second order terms involving derivatives.

It follows from Eq. (5) that if the 4-vector-spinor $\psi_\mu (x)$
satisfies Eq. (9) or Eq. (10), the $\psi_\mu (x)$ also obeys the
Proca Eq. (3), which acts only on vector subspace. But Eq. (3)
involves (this follows after acting the derivative $\partial_\mu$
on the equation) the supplementary condition $\partial_\mu
\varphi_\mu (x)=0$. Thus, Eqs. (9), (10) involve the the
supplementary condition
\begin{equation}
\partial_\mu \psi_\mu (x)=0 . \label{11}
\end{equation}
Eq. (11) reduces the degrees of freedom from $16$ to $12$ because
Eq. (11) is valid for each spinor indexes. So, Eq. (9) (or Eq.
(10)) describes one field with spin-$3/2$ and one field with
spin-$1/2$ (one copy). From the Proca equation and Eq. (11) one
comes to the Klein-Gordon-Fock equation for each component of the
4-vector-spinor
\begin{equation}
 \left(\partial^2_\alpha -m^2\right)\psi_\mu (x)=0 . \label{12}
\end{equation}
Eq. (12) indicates that states with spin $1/2$ and $3/2$ of the
field $\psi_\mu (x)$ have the same mass $m$. Therefore, on the
mass shell we can replace the operator $\partial^2_\alpha$ in Eqs.
(7), (9), (10) by the $m^2$, $\partial^2_\alpha\rightarrow m^2$.
Thus, we arrive from Eqs. (9), (7) to
\begin{equation}
\left(\gamma_\nu\partial_\nu+ m\right)\psi_\mu (x)-
\frac{b}{m}\partial_\mu\partial_\nu \psi_\nu (x)=0 ,
 \label{13}
\end{equation}
where $b=1\pm\sqrt{2}$. The same procedure can be applied to Eq.
(10) which is different from Eq. (9) only by signs in second and
third terms. It should be noted that Eq. (9) and Eq. (13) are not
equivalent each other. Indeed, the supplementary condition (11)
follows from Eq. (9), and Eq. (12) is valid. This means that
non-local Eq. (9) describes particles with spin 3/2 and 1/2 (one
copy) with the unique mass $m$. But this is not the case for Eq.
(13). Eqs. (11), (12) are not consequences of Eq. (13). Therefore,
Eq. (13) describes particles with spin 3/2 and 1/2 (two copies)
with some masses. We obtain the mass spectrum of particles, which
are described by Eq. (13), in the next section. Eq. (13) is
different from the Rarita - Schwinger equation involving two
supplementary conditions \cite{Rarita} which describes pure
spin-3/2 field. In further, we consider the case with arbitrary
parameter $b$ in Eq. (13).

\section{Mass Projection Operators}

The matrix form of Eq. (13) in momentum space for positive energy
is given by
\begin{equation}
\Lambda \psi(p)=0 ,~~\Lambda =i\widehat{p}+m+\frac{b}{m}(p\cdot p)
,
 \label{14}
\end{equation}
where $\widehat{p}=\gamma_\mu p_\mu$, and $(p\cdot p)$ is the
matrix-dyad. For negative energy one should make the replacement
$p\rightarrow -p$.

Acting on Eq. (14) by the operator $i\widehat{p}-m-b(p\cdot p)/m$,
one obtains the equation defining the mass spectrum
\begin{equation}
\left[p^2+m^2+\left(\frac{b^2}{m^2}p^2+2b\right)(p\cdot
p)\right]\psi(p)=0 .
 \label{15}
\end{equation}
The mass matrix
\begin{equation}
M=p^2+m^2+\left(\frac{b^2}{m^2}p^2+2b\right)(p\cdot p)  \label{16}
\end{equation}
obeys the minimal equation
\begin{equation}
\left(M-p^2-m^2\right)\left[M-p^2\left(1+\frac{b^2}{m^2}p^2
+2b\right)-m^2\right]=0 . \label{17}
\end{equation}
Eigenvalues of the matrix $M$ obtained from Eq. (17) are
$\lambda_1= p^2+m^2$, $\lambda_2= p^2\left(1+b^2p^2/m^2
+2b\right)+m^2$. Nontrivial solutions of Eq. (15) exist if
$\mbox{det}M=0$, or equivalently, $\lambda_1=0$ or/and
$\lambda_2=0$. The first solution $\lambda_1=0$ ($p^2=-m^2$)
defines the mass $m$ of the field states. The second solution
$\lambda_2=0$ gives the equation
\begin{equation}
\left[b^2p^4+m^2\left(2b+1\right)p^2+m^4\right]\psi(p)=0 .
 \label{18}
\end{equation}
Eq. (18) possesses solutions:
\begin{equation}
p^2=-m^2\left(\frac{2b+1 \pm \sqrt{4b+1}}{2b^2}\right) ,
 \label{19}
\end{equation}
which define, in general, two different squared masses of fermions
(see \cite{KruglovIJM}, \cite{Kruglov2001} for the case of
bosons):
\begin{equation}
m_1^2=m^2\left(\frac{2b+1 - \sqrt{4b+1}}{2b^2}\right)
,~~m_2^2=m^2\left(\frac{2b+1 + \sqrt{4b+1}}{2b^2}\right) .
 \label{20}
\end{equation}
It follows from Eq. (20) that $b\geq -1/4$. Therefore, the value
$b=1-\sqrt{2}$ does not satisfy this requirement. Both masses are
equal, $m_1=m_2$, at $b=-1/4$. We obtain from Eq. (20) masses of
particles
\begin{equation}
m_1=m\left(\frac{1 - \sqrt{4b+1}}{2b}\right) ,~~m_2=m\left(\frac{1
+\sqrt{4b+1}}{2b}\right) .
 \label{21}
\end{equation}
There are also solutions with opposite signs of masses. From Eqs.
(21), we find the mass formula $m_1+m_2 =m/b$.

Now we consider the projection operators extracting solutions of
Eq. (14) corresponding to the mass $m$ of fields. Straightforward
calculations show that the operator of Eq. (14) obeys the minimal
equation for the value $b=1+\sqrt{2}$:
\begin{equation}
\Lambda^4 -2(1-\sqrt{2})m\Lambda^3
+(1-4\sqrt{2})m^2\Lambda^2-2m^3\Lambda =0 .
 \label{22}
\end{equation}
One may obtain the minimal equation of the operator $\Lambda$ for
arbitrary parameter $b$. Eq. (22) allows us to construct
projection operator extracting a solution of Eq. (14)
corresponding to positive energy. With the help of the method
\cite{Fedorov}, we find such projection operator:
\[
\Pi= -\frac{1}{2m^3}\Lambda^3
+\frac{(1-\sqrt{2})}{m^2}\Lambda^2-\frac{(1-4\sqrt{2})}{2m}\Lambda
+1
\]
\vspace{-7mm}
\begin{equation} \label{23}
\end{equation}
\vspace{-7mm}
\[
=\frac{m-i\widehat{p}}{2m}-\frac{i\widehat{p}(p\cdot
p)}{2m^3}+\frac{(p\cdot p)}{2m^2}
\]
which obeys the necessary equation
\begin{equation}
\Pi^2= \Pi .
 \label{24}
\end{equation}
For negative energy, corresponding to antiparticles, one has to
make the replacement $p\rightarrow -p$ in the operator $\Lambda$.
Every column of the matrix $\Pi$ is the solution, $\psi (p)$, of
Eq. (14). The matrix $\Pi$ has eigenvalues one and zero, so that,
using the transformation $\Pi'=S\Pi S^{-1}$ (the wave function in
this representation becomes $\psi'(p)=S\psi(p)$), the operator
$\Pi$ may be represented in the diagonal form containing only ones
and zeroes. The $\Pi$ acting on the arbitrary non-zero
4-vector-spinor $\chi=\{\chi_\mu\}$ will produce the solution of
Eq. (14), $\psi (p)=\Pi\chi$.

\section{Spin Projection Operators}

The 4-vector-spinor $\psi_\mu$ transforms as a direct product of
the vector representation $(1/2,1/2)$, and the spinor
representation $(1/2,0)\oplus (0,1/2)$ of the Lorentz group. The
generators of the representation $(1/2,1/2)$ can be taken as
follows \cite{Kruglov2001}:
\begin{equation}
J_{\mu\nu}^{(1)}= \varepsilon^{\mu,\nu}-\varepsilon^{\nu,\mu} ,
 \label{25}
\end{equation}
where $4\times 4$-matrices $\varepsilon ^{\mu,\nu}$ obey equations
\cite{Bogush}
\begin{equation}
\left( \varepsilon ^{\mu,\nu}\right) _{\alpha \beta}=\delta
_{\mu\alpha}\delta _{\nu\beta}, \hspace{0.5in}\varepsilon
^{\mu,\sigma}\varepsilon ^{\rho,\nu}=\delta
_{\sigma\rho}\varepsilon ^{\mu,\nu}, \label{26}
\end{equation}
$\mu, \nu=1,2,3,4$. With the help of Eqs. (26) it is easy to
verify that the generators (25) satisfy the commutation relations:
\begin{equation}
\left[ J_{\mu \nu }^{(1)},J_{\alpha \beta}^{(1)}\right] =\delta
_{\nu \alpha }J_{\mu \beta}^{(1)}+\delta _{\mu \beta }J_{\nu
\alpha}^{(1)}-\delta _{\nu \beta }J_{\mu \alpha}^{(1)}-\delta
_{\mu \alpha }J_{\nu \beta}^{(1)} . \label{27}
\end{equation}
It should be noted that in our metrics the antisymmetric
parameters of the Lorentz group $\omega_{mn}$ ($m,n=1,2,3$) are
real, and $\omega_{m4}$ are imaginary.

The generators of the Lorentz group corresponding to the spinor
representation $(1/2,0)\oplus (0,1/2)$ are \cite{Ahieser}
\begin{equation}
J_{\mu\nu}^{(1/2)}= \frac{1}{4}\left( \gamma_\mu
\gamma_\nu-\gamma_\nu \gamma_\mu \right) ,
 \label{28}
\end{equation}
satisfying the algebra (27).

 The generators of the Lorentz group in the
$16$-dimensional representation $(1/2,1/2)\otimes
\left[(1/2,0)\oplus (0,1/2)\right]$ are given by
\begin{equation}
J_{\mu \nu }=J_{\mu \nu }^{(1)}\otimes I_4+I_4 \otimes
J_{\mu\nu}^{(1/2)} , \label{29}
\end{equation}
where the $I_4$ is the unit matrix in 4-dimensional subspaces,
$\otimes$ is the direct product, and $J_{\mu \nu }$ obeys the
algebra (27) corresponding to the Lorentz group.

Let us consider the squared Pauli-Lubanski vector for a vector
subspace
\begin{equation}
\sigma^{(1)2}=\left(\frac{1}{2m}\varepsilon_{\mu\nu\alpha\beta}
p_\nu J_{\alpha\beta}^{(1)}\right)^2
=\frac{1}{m^2}\left(\frac{1}{2}J_{\alpha\beta}^{(1)2} p^2
-J_{\mu\alpha}^{(1)}J_{\nu\alpha}^{(1)}p_\mu p_\nu \right) ,
\label{30}
\end{equation}
where $p^2\equiv p_\mu^2$, $\varepsilon_{\mu\nu\alpha\beta}$ is
antisymmetric tensor Levi-Civita, $\varepsilon_{1234} =-i$. With
the help of Eqs. (25), (26), one may verify that the operator (30)
obeys the minimal matrix equation
\begin{equation}
\sigma^{(1)2}\left(\sigma^{(1)2}-2\right)=0 . \label{31}
\end{equation}
The eigenvalues of the squared spin operator $\sigma^{(1)2}$ equal
to $s(s+1)$. Eq. (31) shows that the eigenvalues of
$\sigma^{(1)2}$ are zero and two, corresponding to spins $s=0$ and
$s=1$. This is the consequence of the fact that the representation
$(1/2,1/2)$ of the Lorentz group includes two spins, $s=0,1$ (see
\cite{Kruglov}, \cite{Kruglov2001}).

The squared Pauli-Lubanski vector for a spinor subspace
$\sigma^{(1/2)2}=\left(\varepsilon_{\mu\nu\alpha\beta} p_\nu
J_{\alpha\beta}^{(1/2)}/2m\right)^2$ becomes
\begin{equation}
\sigma^{(1/2)2}=\frac{3}{4} , \label{32}
\end{equation}
indicating that the representation $(1/2,0)\oplus (0,1/2)$
involves only pure spin $1/2$.

The $16$-dimensional representation $(1/2,1/2)\otimes \left[
(1/2,0)\oplus (0,1/2)\right]$ considered can be represented as
$(1,1/2)\oplus (1/2,1)\oplus(1/2,0)\oplus (0,1/2)$ and contains
multi-spin $1/2, 3/2$. Indeed, one can verify that the squared
Pauli-Lubanski vector for this representation
\[
\sigma^{2}=\frac{1}{m^2}\left(\frac{1}{2}J_{\alpha\beta}^{2} p^2
-J_{\mu\alpha}J_{\nu\alpha}p_\mu p_\nu \right) ,
\]
with the generators (29) satisfies the minimal matrix equation
\begin{equation}
\left(\sigma^{2} -
\frac{3}{4}\right)\left(\sigma^{2}-\frac{15}{4}\right)=0 .
\label{33}
\end{equation}
As a result, the equation
\begin{equation}
\sigma^{2}\psi (p)=s(s+1)\psi (p)  \label{34}
\end{equation}
corresponds to eigenvalues $s(s+1)=3/4$ ($s=1/2$) and
$s(s+1)=15/4$ ($s=3/2$). To separate states with spin $s=1/2$ and
$3/2$, we explore the the method described in \cite{Fedorov}.
Following this procedure one arrives at the projection operators
\begin{equation}
S^{2}_{1/2}=\frac{5}{4}-\frac{1}{3}\sigma^2 ,~~~~
S^{2}_{3/2}=\frac{1}{3}\sigma^2-\frac{1}{4} . \label{35}
\end{equation}
The operators $S^{2}_{1/2}$, $S^{2}_{3/2}$ acting on the
16-dimensional vector extract states with spin $1/2$ and $3/2$,
correspondingly, obey equations similar to Eq. (24) and also the
equations: $S^{2}_{1/2}+S^{2}_{3/2}=1$ ($1\equiv I_{16}$ is a unit
matrix in 16-dimensional space), $S^{2}_{1/2}S^{2}_{3/2}=0$.

The total set of commuting operators includes also operators of
the spin projections on the direction of the momentum
$\textbf{p}$. Thus, we introduce spin projection operators
\begin{equation}
\sigma_p=-\frac{i}{2|\textbf{p}|}\epsilon_{abc}\textbf{p}_a
J_{bc}=\sigma_p^{(1)}\otimes I_4 +I_4\otimes\sigma_p^{(1/2)} ,
\label{36}
\end{equation}
where $|\textbf{p}| =\sqrt{p_1^2 +p_2^2+p_3^2}$, and spin
projection operators for vector and spinor subspaces are defined
as
\begin{equation}
\sigma_p^{(1)}=-\frac{i}{2|\textbf{p}|}\epsilon_{abc}\textbf{p}_a
J_{bc}^{(1)}
,~~~~\sigma_p^{(1/2)}-\frac{i}{2|\textbf{p}|}\epsilon_{abc}\textbf{p}_a
J_{bc}^{(1/2)} . \label{37}
\end{equation}
Using Eqs. (6), (26), one may verify that minimal matrix equations
\begin{equation}
\sigma_p^{(1)}\left(\sigma_p^{(1)}-1\right)\left(\sigma_p^{(1)}+1\right)=0
,~~~~\left(\sigma_p^{(1/2)}-\frac{1}{2}\right)\left(\sigma_p^{(1/2)}+\frac{1}{2}\right)=0
, \label{38}
\end{equation}
\begin{equation}
\left(\sigma_p^2-\frac{1}{4}\right)\left(\sigma_p^2-\frac{9}{4}\right)=0
, \label{39}
\end{equation}
are valid. With the aid of method \cite{Fedorov}, we construct
from Eq. (39) the projection operators
\begin{equation}
P_{\pm 1/2}=\mp\frac{1}{2}\left(\sigma_p\pm\frac{1}{2}
\right)\left(\sigma_p^2-\frac{9}{4}\right) ,~~~~ P_{\pm
3/2}=\pm\frac{1}{6}\left(\sigma_p\pm\frac{3}{2}
\right)\left(\sigma_p^2-\frac{1}{4}\right) \label{40}
\end{equation}
which extract spin projections $\pm 1/2$, $\pm 3/2$. These
operators obey the required properties (24) and
\begin{equation}
P_{+1/2}+P_{-1/2}+P_{+ 3/2}+P_{- 3/2}=1 ,~~~~P_{\pm 1/2}P_{\pm
3/2}=0 . \label{41}
\end{equation}

The projection operators extracting states with pure spin, spin
projections and positive energy are given by
\begin{equation}
\Delta^{(1/2)}_{\pm 1/2}= \Pi S^{2}_{1/2} P_{\pm 1/2} , \label{42}
\end{equation}
\begin{equation}
\Delta^{(3/2)}_{\pm 1/2}= \Pi S^{2}_{3/2} P_{\pm 1/2} ,~~~~
\Delta^{(3/2)}_{\pm 3/2}= \Pi S^{2}_{3/2} P_{\pm 3/2} . \label{43}
\end{equation}
Thus, the projection matrix $\Delta^{(1/2)}_{\pm 1/2}$ extracts
the states with spin $1/2$ and spin projections $\pm 1/2$, and
$\Delta^{(3/2)}_{\pm 1/2}$, $\Delta^{(3/2)}_{\pm 3/2}$ correspond
to spin $3/2$ and spin projections $\pm 1/2$ and $\pm 3/2$,
correspondingly. The projection operators $\Delta^{(1/2)}_{\pm
1/2}$, $\Delta^{(3/2)}_{\pm 1/2}$, $\Delta^{(3/2)}_{\pm 3/2}$ are
also the density matrices for pure spin states. To obtain
projection matrices for negative energy, one has to make the
replacement $p\rightarrow -p$ in the mass operator $\Lambda$, Eqs.
(14), (23). We may get impure states of fields by summation of
Eqs. (42), (43) over spin projections and spins.

\section{First Order Relativistic wave Equation}

To formulate the second order (in derivatives) Eq. (14) in the
form of the first order relativistic wave equation, we introduce
the 20-dimensional function
\begin{equation}
\Psi (x)=\left\{ \psi _A(x)\right\} =\left(
\begin{array}{c}
\psi _0 (x)\\
\psi _\mu (x)
\end{array}
\right) \hspace{0.5in}(\psi_0 (x)=-\frac{1}{m}\partial_\mu
\psi_\mu (x)) , \label{44}
\end{equation}
where $A=0,\mu$; $\psi_0 (x)$ is bispinor. With the aid of the
elements of the entire algebra Eq. (26), Eq. (13) may be written
in the form of one equation
\begin{equation}
\partial _\nu \left(\varepsilon ^{0,\nu }+ b\varepsilon ^{\nu ,0} +
\varepsilon^{\alpha,\alpha}\gamma_\nu\right)_{AB}\Psi _B(x)+
m\left[ \varepsilon ^{\mu ,\mu }+ \varepsilon ^{0,0}\right]
_{AB}\Psi _B(x)=0 , \label{45}
\end{equation}
where $\gamma$-matrices act on spinor indexes. Introducing
20-dimensional matrices
\[
\beta _\nu =\left(\varepsilon ^{0,\nu }+ b\varepsilon ^{\nu
,0}\right)\otimes I_4 +
\varepsilon^{\alpha,\alpha}\otimes\gamma_\nu,
\]
\begin{equation}
1\equiv I_{20}= I_5\otimes I_4 ,~~~~I_5=\varepsilon ^{\mu ,\mu
}+\varepsilon ^{0,0} ,\label{46}
\end{equation}
where $I_{20}$ is the unit matrix in 20-dimensional space, Eq.
(45) becomes the relativistic wave equation of the first order:
\begin{equation}
\left( \beta _\mu \partial _\mu +m\right) \Psi (x)=0 . \label{47}
\end{equation}
It should be noted that the unit matrix $I_4\equiv \varepsilon
^{\alpha ,\alpha }$ (we imply a summation over repeating indexes)
in Eq. (46) acts in the 4-dimensional vector subspace. The
generators of the Lorentz group in 20-dimensional space
\begin{equation}
J_{\mu \nu }=J_{\mu \nu }^{(1)}\otimes I_4+I_5 \otimes
J_{\mu\nu}^{(1/2)} , \label{48}
\end{equation}
obey the commutation relations (27) and the commutation relations
\begin{equation}
\left[ \beta _\lambda ,J_{\mu \nu }\right] =\delta _{\lambda \mu
}\beta _\nu -\delta _{\lambda \nu }\beta _\mu .  \label{49}
\end{equation}

Eq. (49) guarantees the form-invariance of Eq. (47) under the
Lorentz transformations.

The Hermitianizing matrix $\eta$ for a 20-dimensional field
$\Psi_\mu$ is given by
\begin{equation}
\eta=\left(\varepsilon^{m,m}-\varepsilon^{4,4}-b\varepsilon^{0,0}\right)\otimes
\gamma_4 , \label{50}
\end{equation}
and obeys the necessary equations \cite{Bogush}, \cite{Fedorov}:
\begin{equation}
\eta \beta _i=-\beta _î^{+}\eta ,\hspace{0.5in}\eta \beta _4=\beta
_4^{+}\eta \hspace{0.5in}(i=1,2,3) .  \label{51}
\end{equation}
Thus, that a relativistically invariant bilinear form is
\begin{equation}
\overline{\Psi }\Psi =\Psi ^{+}\eta \Psi ,  \label{52}
\end{equation}
where $\Psi ^{+}$ is the Hermitian-conjugate wave function. Eq.
(52) allows us to consider the Lagrangian in the standard form (up
to a total derivative):
\begin{equation}
{\cal L}=-\overline{\Psi }(x)\left(\beta _\mu \partial _\mu
+m\right) \Psi (x) .  \label{53}
\end{equation}
The variation of the action corresponding to the Lagrangian (53)
produces the Euler-Lagrange equation (47).

In accordance with the general theory \cite{Gel'fand}, Eq. (47)
defines the masses of particles: $m/\theta_i$, where $\theta_i$
being the eigenvalues of the matrix $\beta_4$. The matrix $\beta
_4 =\left(\varepsilon ^{0,4 }+ b\varepsilon ^{4 ,0}\right)\otimes
I_4 + \varepsilon^{\alpha,\alpha}\otimes\gamma_4$ obeys the
minimal matrix equation
\begin{equation}
\left(\beta_4^2 -1\right)\left[\beta_4^4
-\left(2b+1\right)\beta_4^2 +b^2\right] =0 . \label{54}
\end{equation}
We obtain from Eq. (54) six eigenvalues of the matrix $\beta_4$:
$\pm 1$, $\pm\theta_1$, $\pm\theta_2$, where
\begin{equation}
\theta_1=\frac{-1 - \sqrt{4b+1}}{2},~~\theta_2=\frac{-1 +
\sqrt{4b+1}}{2} . \label{55}
\end{equation}
It is easy to prove that field masses: $m_1=m/\theta_1$,
$m_2=m/\theta_2$ coincide with expressions in Eq. (21).

It should be noted that Eqs. (47)-(55) are valid for arbitrary
parameter $b$. It is not difficult also to obtain the spin
projection operators in 20-dimensional representation space using
the technique of Sec. 4.

\section{Electromagnetic Interactions of Fields}

As follows from Eq. (54), the inverse matrix $\beta_4^{-1}$
exists, and is given by
\begin{equation}
\beta_4^{-1}=\frac{1}{b^2}\left[\beta_4^5-\left(2b+2\right)\beta_4^3
+ (b+1)^2\beta_4\right] . \label{56}
\end{equation}
Therefore, all components of the wave function $\Psi (x)$ contain
time derivatives and no subsidiary conditions in the first order
wave equation (47). The minimal interaction with electromagnetic
fields can be obtained by the substitution $\partial_\mu
\rightarrow D_\mu = \partial_\mu -ieA_\mu$ ($A_\mu$ is the
4-vector-potential of the electromagnetic field) in Eqs. (13),
(47), (53). We also introduce non-minimal interaction in Eq. (47)
as follows:
\begin{equation}
\biggl [\beta _\mu D_\mu +\frac i2\left( \sigma _0P_0+\sigma _1
P_1\right) \beta _{\mu \nu }\mathcal{F}_{\mu \nu }+m\biggr ]\Psi
(x)=0  ,\label{57}
\end{equation}
where $\mathcal{F}_{\mu \nu }=\partial _\mu A_\nu -\partial _\nu
A_\mu $ is the strength of the electromagnetic field,
\[
\beta _{\mu \nu }=\beta _\mu \beta _\nu -\beta _\nu \beta _\mu
\]
\begin{equation}
=I_4\otimes\left(\gamma_\mu\gamma_\nu-\gamma_\nu\gamma_\mu\right)
+b\left(\varepsilon^{\nu,0}\otimes\gamma_\mu
-\varepsilon^{\mu,0}\otimes \gamma_\nu\right)
 \label{58}
\end{equation}
\[
+\varepsilon^{0,\mu}\otimes\gamma_\nu-\varepsilon^{0,\nu}
\otimes\gamma_\mu +b \left(\varepsilon^{\mu,\nu} -
\varepsilon^{\nu,\mu} \right)\otimes I_4.
\]
and projection operators $P_0$, $P_1$, are
\begin{equation}
P_0=\varepsilon ^{0,0}\otimes I_4,\hspace{0.3in}P_1=\varepsilon
^{\mu ,\mu }\otimes I_4 .
 \label{59}
\end{equation}
The projection operators $P_0$, $P_1$ obey the relations:
$P_0^2=P_0$, $P_1^2=P_1$, $P_0+P_1=1$. Eq. (57) is form-invariant
under the Lorentz transformations and contains additional
parameters $\sigma_0$, $\sigma_1$ characterizing anomalous
electromagnetic interactions of fermions.

Using Eqs. (26), (44), (46), from Eq. (57), one may obtain
equations:
\[
\left( \gamma_\nu D_\nu +i\sigma_1 \gamma_\mu \gamma_\nu
\mathcal{F}_{\mu \nu} + m\right)\psi_\mu (x)+b\left(D_\mu
+i\sigma_1\gamma_\nu\mathcal{F}_{\nu \mu}\right)\psi_0 (x)
\]
\vspace{-7mm}
\begin{equation}
\label{60}
\end{equation}
\vspace{-7mm}
\[
+i\sigma_1 b\mathcal{F}_{\mu \nu}\psi_\nu (x)=0 ,
 \]
\begin{equation}
\left(D _\mu +i\sigma_0\gamma_\nu\mathcal{F}_{\mu \nu}\right)
\psi_\mu (x)+ \left(m +i\sigma_0 \gamma_\mu
\gamma_\nu\mathcal{F}_{\mu \nu}\right)\psi_0 (x)=0 .
 \label{61}
\end{equation}
Expressing the spinor $\psi_0 (x)$ from Eq. (61) and replacing it
into Eq. (60), one may obtain an equation for vector-spinor
$\psi_\mu (x)$. The system of equations (60), (61) describes the
non-minimal electromagnetic interaction of particles with
multi-spin 3/2, 1/2. Possibly, these equations can be used for a
description of interacting composite fermions.

Obviously, one may find the quantum-mechanical Hamiltonian from
Eq. (57). Indeed, we obtain
\[
i\beta _4\partial _t\Psi (x)=\biggl [\beta _aD_a+m+eA_0\beta _4+
\]
\begin{equation}
+\frac{i}{2}\left( \sigma _0 P_0+\sigma _1P_1 \right) \beta _{\mu
\nu }\mathcal{F}_{\mu \nu }\biggr ]\Psi (x).  \label{62}
\end{equation}
Then, the Hamiltonian form of the equation is
\[
i\partial _t\Psi (x)=\mathcal{H}\Psi (x) ,
\]
\vspace{-7mm}
\begin{equation} \label{63}
\end{equation}
\vspace{-7mm}
\[
\mathcal{H}=\beta _4^{-1}\biggl [\beta _aD_a+m+eA_0\beta _4+ \frac
i2\left( \sigma _0P_0+\sigma _1P_1 \right) \beta _{\mu \nu
}\mathcal{F}_{\mu \nu }\biggr ] ,
\]
where $\beta _4$ is given by Eq. (56).

Now we consider the consistency problem. The noncausal propagation
of particles in the Rarita-Schwinger theory is connected with the
subsidiary conditions (constraints). Indeed, due to these
conditions there are non-dynamical auxiliary components of the
wave function which should be eliminated. But for definite field
strength of the electromagnetic fields it is impossible. As a
result, there are noncausal propagating modes in the
Rarita-Schwinger approach. We investigate this question for Eq.
(57). To clear up the number of dynamical variables of the wave
function (44), one has to consider the matrix $\beta_4$ in Eq.
(57). As there exists the inverse matrix $\beta_4^{-1}$, all
components of the wave function (44) are canonical and there are
no subsidiary conditions in the first order wave equation (57).
According to the method \cite{Zwanziger}, one should replace the
derivatives in Eq. (57) by the four-vector $n_\mu$ to investigate
the characteristic surfaces. We may find the normals to the
characteristic surfaces, $n_\mu$, by considering the equation
\begin{equation}
 \det\left(\beta_\mu n_\mu\right) =0 . \label{64}
\end{equation}
If there are noncausal propagating components, there should be
non-trivial solution to Eq. (64) for the time-like vector $n_\mu$.
In the frame of reference where $n_\mu=(0,0,0,n_4)$, Eq. (64)
becomes $\det(\beta_4 n_4)=0$. As there are no zero eigenvalues of
the matrix $\beta_4$ due to Eq. (54), equation $\det(\beta_4
n_4)=0$ has only the trivial solution $n_4=0$, and no superluminal
speed of waves in the approach considered. This is a consequence
that Eq. (57) describes particles with multi-spin 1/2, 3/2. One
can see the possible applications of multi-spin approach to the
construction of gauge theories based on spin degrees of freedom
and hadron phenomenology in \cite{Kruglov2001}.

\section{Conclusion}

We have shown that the non-local equation (9) suggested describes
the field with spin 3/2 and 1/2 (one copy) and the unique mass
$m$, and is related in a direct manner with the Proca equation.
This equation involves only one supplementary condition
$\partial_\mu\psi_\mu=0$, which removes one state with spin-1/2
but remains another state with spin-1/2. We show that suggested
local Eq. (13) describes particles with spin 3/2 and 1/2 (two
copies) and three different masses: $m$, $m_1$, $m_2$, so that
$m_1 +m_2 =m/b$. The massless particles obey nonlocal Eqs. (9),
(10) at $m=0$, $a=\sqrt{\partial_\mu^2}$.

The projection operators extracting solutions of the field
equation with positive and negative energy are constructed. We
obtain spin projection operators which allow us to separate states
with spins $1/2$ and $3/2$ and spin projections. The first order
relativistic wave equation in the 20-dimensional matrix form is
formulated and the relativistically invariant bilinear form and
the corresponding Lagrangian are given. We have introduced two
parameters characterizing non-minimal electromagnetic interactions
of fermions and have found the quantum-mechanical Hamiltonian. It
was proved that there is only causal propagation of particles in
the approach considered and the theory is consistent.

The question about the gyromagnetic ratio $g$ was unanswered here.
This is important for unitarity requirement and phenomenological
applications. It is known \cite{Weinberg} that low energy
unitarity requires the gyromagnetic ratio to be $g=2$ for
arbitrary spin. We leave this question for further investigations.

Some possible applications of the equations studied: the effective
theory of hadron interactions and the description of massive
gravitinos in the theory of supergravity. This, however, requires
further investigations of the equations suggested.

\end{document}